\documentclass{JINST}
\usepackage{times}
\usepackage{cite,mcite}
\usepackage{graphicx}
\usepackage{subfigure}
\usepackage{atlasphysics}
\usepackage{mathptmx}
\usepackage{url}
\newcommand{\pib}{\mbox{\boldmath$\pi$}}
\newcommand{\degree}{\ensuremath{^\circ}}
\title{Alignment of the Pixel and SCT Modules for the 2004 ATLAS Combined Test Beam}
\author{
 A.~Ahmad$^{35}$$^{,a}$,
 A.~Andreazza$^{21}$,
 T.~Atkinson$^{20}$,
 J.~Baines$^{31}$,
 A.J.~Barr$^{28}$,
 R.~Beccherle$^{11}$,
 P.J.~Bell$^{17}$,
 J.~Bernabeu$^{38}$,
 Z.~Broklova$^{30}$,
 P. A.~Bruckman de Renstrom$^{28}$$^{,b}$
% ...
\\
 \llap{$^{1}$} University of Bergen, Department for Physics and Technology, Allegaten 55, NO - 5007 Bergen, Norway\\
\llap{$^{2}$} Lawrence Berkeley National Laboratory and University of California, Physics Division, MS50B-6227, 1 Cyclotron Road, Berkeley, CA 94720, United States of America\\
\llap{$^{3}$} Physikalisches Institut der Universitaet Bonn, Nussallee 12, D - 53115 Bonn, Germany\\
\llap{$^{4}$} Brookhaven National Laboratory, Physics Department, Bldg. 510A, Upton, NY 11973, United States of America\\
\llap{$^{5}$} Cavendish Laboratory, University of Cambridge, J J Thomson Avenue, Cambridge CB3 0HE, United Kingdom\\
\llap{$^{6}$} CERN, CH - 1211 Geneva 23, Switzerland\\
\llap{$^{7}$} Niels Bohr Institute, University of Copenhagen, Blegdamsvej 17, DK - 2100 Kobenhavn 0, Denmark\\
\llap{$^{8}$} Faculty of Physics and Applied Computer Science of the AGH-University of Science and Technology, (FPACS, AGH-UST), al. Mickiewicza 30, PL-30059 Cracow, Poland\\
\llap{$^{9}$} Universitaet Dortmund, Experimentelle Physik IV, DE - 44221 Dortmund, Germany\\
 \llap{$^{10}$} Universite de Geneve, Section de Physique, 24 rue Ernest Ansermet, CH - 1211 Geneve 4, Switzerland\\
 \llap{$^{11}$} INFN Genova and Universit\`a  di Genova, Dipartimento di Fisica, via Dodecaneso 33, IT - 16146 Genova, Italy\\
 \llap{$^{12}$} University of Glasgow, Department of Physics and Astronomy, UK - Glasgow G12 8QQ, United Kingdom\\
 \llap{$^{13}$} Hampton University, Department of Physics, Hampton, VA 23668, United States of America\\
 \llap{$^{14}$} Institut fuer Astro- und Teilchenphysik, Technikerstrasse 25, A - 6020 Innsbruck, Austria\\
 \llap{$^{15}$} Joint Institute for Nuclear Research, JINR Dubna, RU - 141 980 Moscow Region, Russia\\
 \llap{$^{16}$} University College London, Department of Physics and Astronomy, Gower Street, London WC1E 6BT, United Kingdom\\
 \llap{$^{17}$} School of Physics and Astronomy,  University of Manchester, UK - Manchester M13 9PL, United Kingdom\\
 \llap{$^{18}$} Universitaet Mannheim, Lehrstuhl fuer Informatik V, B6, 23-29, DE - 68131 Mannheim, Germany \\
 \llap{$^{19}$} CPPM, Aix-Marseille Universit, CNRS/IN2P3, Marseille, France\\
 \llap{$^{20}$} School of Physics, University of Melbourne, AU - Parkvill, Victoria 3010, Australia\\
 \llap{$^{21}$} INFN Milano and Universit\`a  di Milano, Dipartimento di Fisica, via Celoria 16, IT - 20133 Milano, Italy\\
 \llap{$^{22}$} P.N. Lebedev Institute of Physics, Academy of Sciences, Leninsky pr. 53, RU - 117 924 Moscow, Russia\\
 \llap{$^{23}$} Moscow Engineering \& Physics Institute (MEPhI), Kashirskoe Shosse 31, RU - 115409 Moscow, Russia\\
 \llap{$^{24}$} Max-Planck-Institut f\"ur Physik, (Werner-Heisenberg-Institut), F\"ohringer Ring 6, 80805 M\"unchen, Germany\\
 \llap{$^{25}$} Radboud University Nijmegen/NIKHEF, Dept. of Exp. High Energy Physics, Toernooiveld 1, NL - 6525 ED Nijmegen , Netherlands \\
 \llap{$^{26}$} Nikhef National Institute for Subatomic Physics, Kruislaan 409, P.O. Box 41882, NL - 1009 DB Amsterdam, Netherlands\\
 \llap{$^{27}$} Ohio State University, 191 West Woodruff Ave, Columbus, OH 43210-1117, United States of America\\
 \llap{$^{28}$} Department of Physics, Oxford University, Denys Wilkinson Building, Keble Road, Oxford OX1 3RH, United Kingdom\\
 \llap{$^{29}$} Laboratorio de Instrumentacao e Fisica Experimental de Particulas - LIP, and SIM/Univ. de Lisboa, Avenida Elias Garcia 14-1, PT - 1000-149, Lisboa, Portugal \\
}\author{{}\\
 \llap{$^{30}$} Charles University in Prague, Faculty of Mathematics and Physics, Institute of Particle and Nuclear Physics, V Holesovickach 2, CZ - 18000 Praha 8, Czech Republic\\
 \llap{$^{31}$} Rutherford Appleton Laboratory, Science and Technology Facilities Council, Harwell Science and Innovation Campus, Didcot OX11 0QX, United Kingdom\\
 \llap{$^{32}$} CEA, DSM/DAPNIA, Centre d’Etudes de Saclay, FR - 91191 Gif-sur-Yvette, France\\
 \llap{$^{33}$} University of California Santa Cruz, Santa Cruz Institute for Particle Physics (SCIPP), Santa Cruz, CA 95064, United States of America\\
 \llap{$^{34}$} University of Sheffield, Department of Physics \& Astronomy, Hounsfield Road, UK - Sheffield S3 7RH, United Kingdom\\
 \llap{$^{35}$} Insitute of Physics, Academia Sinica, TW - Taipei 11529, Taiwan\\
 \llap{$^{36}$} University of Tsukuba, Institute of Pure and Applied Sciences, 1-1-1 Tennoudai, Tsukuba-shi, JP - Ibaraki 305-8571, Japan\\
 \llap{$^{37}$} INFN Gruppo Collegato di Udine and Universit\`{a} di Udine, Dipartimento di Fisica, via delle Scienze 208, IT - 33100 Udine; INFN Gruppo Collegato di Udine and ICTP, Strada Costiera 11, IT - 34014 Trieste, Italy\\
 \llap{$^{38}$} Instituto de F\'isica Corpuscular (IFIC), Centro Mixto UVEG-CSIC, Apdo. 22085,  ES-46071 Valencia; Dept. F\'isica At., Mol. y Nuclear, Univ. of Valencia and Instituto de Microelectr\'onica de Barcelona (IMB-CNM-CSIC), 08193 Bellaterra, Barcelona, Spain\\
 \llap{$^{39}$} Bergische Universitaet, Fachbereich C, Physik, Postfach 100127, Gauss-Strasse 20, DE-42097 Wuppertal, Germany\\
\\
$^{a}$ Now at State University of New York at Stony Brook \\
$^{b}$ also at H. Niewodniczanski Institute of Nuclear Physics PAN, Cracow, Poland\\

Email: \email{muge.karagoz.unel@cern.ch,martis@ific.uv.es}
}

\abstract{A small set of final prototypes of the ATLAS Inner Detector
silicon tracker (Pixel and SCT) were used to take data during the 2004 Combined
Test Beam.  Data were collected from runs with beams of different
flavour (electrons, pions, muons and photons) with a momentum
range of 2 to 180 GeV/c. Four independent methods were used to
align the silicon modules.  The corrections obtained were validated
using the known momenta of the beam particles and were shown to yield
consistent results among the different alignment approaches.  From the
residual distributions, it is concluded that the precision attained in the
alignment of the silicon modules is of the order of 5~$\mu$m in their
most precise coordinate.
}
\keywords{Detector alignment and calibration methods, Solid state detectors, Particle tracking detectors, 
Large detector systems for particle and astroparticle physics}

\begin{document}
%
\section{Introduction \label{sec:Intro}}
This note reports the results of the alignment of the ATLAS Inner Detector
\cite{IDTDR} silicon tracker (Pixel and  SCT) modules at the ATLAS Combined Test Beam
data-taking (CTB) which took place at the CERN H8 beam-test facility in 2004. The purpose
of the CTB was to study the combined performance of ATLAS. The setup
represented a full barrel slice of the Inner Detector (ID), Calorimeter and Muon
Spectrometer of the complete ATLAS detector and
was instrumented with final prototypes.

Once the Pixel and SCT modules had been installed in the CTB setup in 
addition to the already operational TRT, the Inner Detector 
was fully integrated into the common data acquisition system. 
Data were collected with this fully integrated ID, 
using beams with different characteristics.
Pion, electron, muon and photon beams were used in a wide
range of momenta from 2 to 180 $GeV/c$,  and some data were taken
without magnetic field (B).

The CTB setup represented an ideal framework for testing the Inner
Detector software. The offline reconstruction was tested on real data using
the ATLAS software framework (ATHENA) \cite{atlas_computing_model} and
was particularly useful for tracking \cite{Thijs_phd}, 
pre-commissioning tests, and for testing the alignment software.

Determining the locations of the tracking detector elements is crucial 
for the performance of the ID tracker.
For this purpose, various alignment algorithms, based on optimization of 
track hit residuals, were applied to align the CTB silicon setup.  
An alignment
algorithm specifically developed for the CTB (hereafter referred to as 
\Valencia approach \cite{Sergio_phd}) had been adapted from an 
algorithm used in previous SCT standalone test beams
\cite{sct_testbeam}, by the time the first data were collected. The \Valencia approach
produced alignment corrections  for the initial
CTB data analysis.  For the final  analysis of the alignment,  three more
algorithms were tested. These algorithms, developed for the alignment
of the entire  Inner Detector silicon tracker, are: 
\Robust \cite{robustalignalg}, 
\Local \cite{MPP-2005-174,MPP-2006-118} and 
\Global \cite{globalX2_note,globalX2_mumbai} 
approaches \cite{AlignmentYellowBook}.

The resulting sets of alignment constants were used to measure the momenta
of the incident particles in electron and pion runs. A comparison with
the nominal momenta was used to crosscheck the different
alignment procedures.  The residual distributions and 
reconstructed track parameters were studied for electrons and pions
with and without B field. The global reference frame was also studied by matching the 
alignment results via a global offset optimization.

%
%
\section{Setup, Data Samples and Tracking  \label{sec:setup_data}}
%
%
The Inner Detector volume in the CTB setup was divided into three containers for
each sub-detector: Pixel, SCT and TRT. Six Pixel and eight SCT modules were placed in their 
respective containers\footnote{The ATLAS detector has, in total, 1744 Pixel modules and 4088 SCT modules.}. 
The TRT setup consisted of two barrel wedges,
equivalent to 1/16 of the circumference of a cylinder.

\begin{figure}[htb]
\begin{center}
\begin{minipage}{.48\textwidth}
\centering
\includegraphics[width=\textwidth]{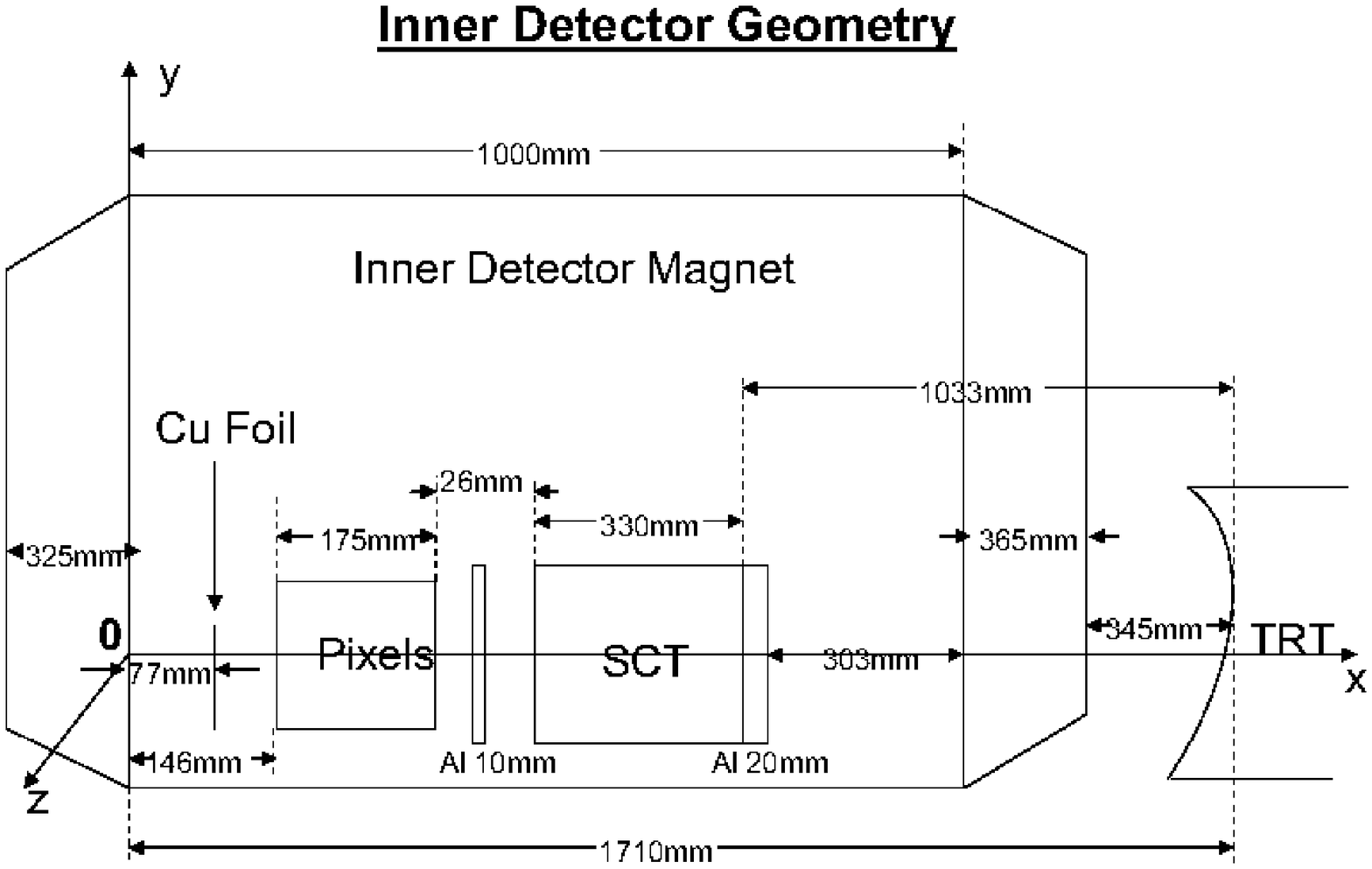}
\end{minipage}
\hspace{1cm}
\begin{minipage}{.4\linewidth}
\centering
\includegraphics[width=\textwidth]{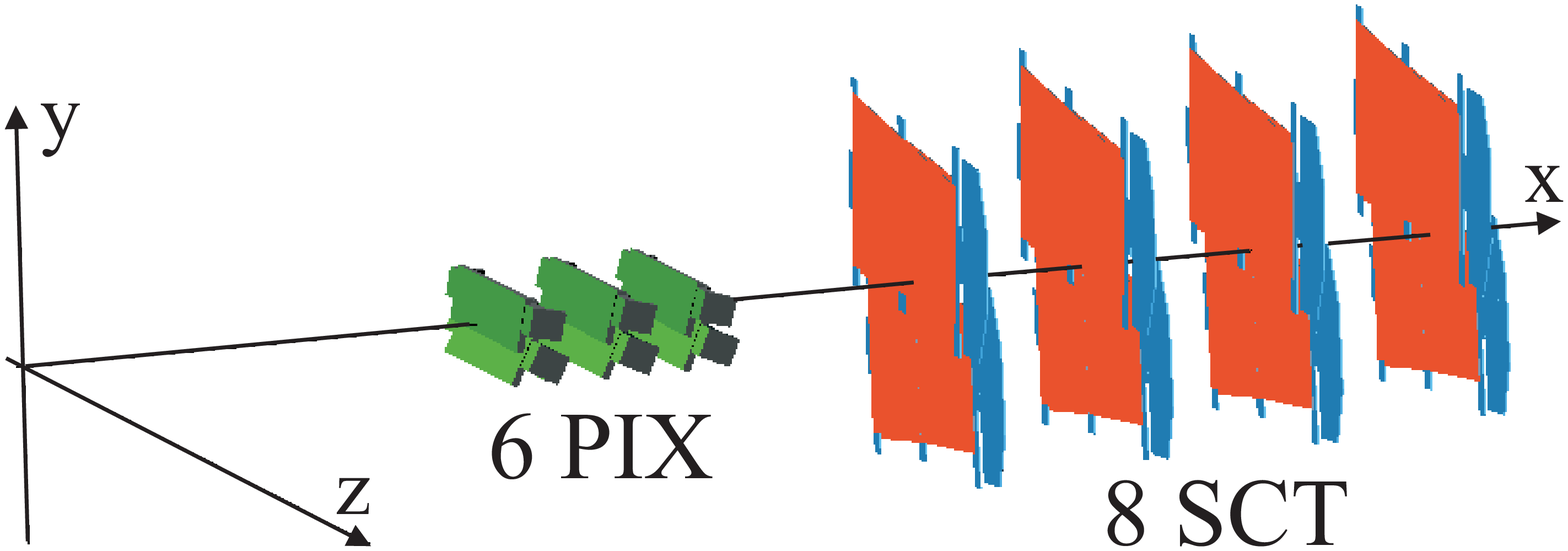}
\end{minipage}
\caption{Schematic representation of the ID components 
and the magnet at the CTB. The reference coordinate 
system is also shown. The long pixel coordinate and the SCT module 
strips are almost parallel to the $Z$-axis.\label{fig:idcomp}}
\end{center}
\end{figure}

The coordinate system was chosen
to be right-handed, with the $X$-axis along the beam direction and the
$Y$-axis pointing vertically upward as depicted in
Fig.~\ref{fig:idcomp} \cite{CTB}. The origin was located at the
entrance of the dipole magnet that produced a maximum 1.4 T field
in the negative Z-direction.  
The Pixel and SCT detectors
were located inside the magnet whereas the TRT detector was located outside
due to its larger dimension.

A Pixel module \cite{pixel_sensors,pixel_vertex2001} consists of a
single silicon wafer with an array of 50$\times$400 $\mu$m$^2$ pixels that
are read out by 16 chips \cite{pixel_chip}. 
The active area of each module is $\sim 60.8 \times 16.4 \mm^2$.
In the CTB setup, six Pixel modules were distributed in three layers (0,1,2) and two
sectors (0,1). 
The distances along the beam axis between the different layers and
the locations of modules within each layer mimic the arrangement
of the modules in ATLAS.
The first Pixel layer was nominally located at 195.986~mm from 
the global coordinate center along the 
beamline ($X$-axis) and the last layer was located at 268.277~mm.
Each module was positioned at an
angle of about $20\degree$ with respect to the incident beam, around the long pixel coordinate.
Modules in the same layer overlapped by $\sim 200~\mu$m.

A SCT module is built from four single-sided silicon microstrip sensors
glued back to back in pairs with 40~mrad stereo
angle for a 3D space-point reconstruction \cite{SCT_sensor,sct_endcap_module}.
The modules produce two hits, one in each plane. 
The SCT end-cap modules have a wedge-shaped geometry 
which results in variable pitch sizes (Fig.~\ref{fig:angular_res}). 
In the CTB setup, one of the four shape-wise distinct SCT end-cap module types 
was used (outer module). 
For the outer end-cap modules, 
the readout strip pitch is 70.9-81.1~$\mu$m.
Each plane has a length of about 120.0 $\mm$ and bases of about $72.0~\mm$ and $57.0~\mm$. 
The readout is provided by a binary chip \cite{sct_abcd_chip}.  
Eight SCT modules were used in each of the four layers (0,3) of the CTB setup; 
distributed in two sectors (0,1) with a $4\mm$ overlap.  
The arrangement of the modules was similar to the SCT barrel configuration in ATLAS\footnote{The rectangular barrel modules
which have uniform $80~\mu m$ pitch were not used due to their unavailability during test beam data-taking.}, 
however, the modules were not mounted at an angle with respect to the beam axis. 
The SCT modules were nominally positioned from 378.198~mm to 598.218~mm along the beam axis.

The beam-line instrumentation, including trigger and veto scintillators,
Cherenkov counters and readout system is documented elsewhere 
\cite{ctb_beam, ctb_daq}. The Inner Detector magnetic
field profile was measured\cite{CTB} and its 
non-uniformity was taken into account during the track reconstruction. 
The absolute momentum as measured by the silicon detector which was 
located in a very uniform magnetic field region was certified to better than 
1$\%$ by comparing the momentum reconstructed from silicon alone with that 
obtained independently using the angular measurement in the TRT.


The CTB ID data taking was divided into five different periods between
September 2004 and November 2004~\cite{CTB}, where 22 million usable 
events were collected. In order to evaluate the
material effects in the tracker, aluminum plates (10\% $X_0$) were
inserted and removed between the Pixel, SCT and TRT
setups (Fig.  \ref{fig:idcomp}) in alternate runs. The TRT was repositioned in the
transverse plane of the beam.   Particle type and energy of the beam
also alternated during the periods.

The algorithms provided a valid silicon detector alignment for all the CTB data-taking periods.
However, this article reports
on the last period (period 5) of stable data-taking when no extra material layers were
used. Table~\ref{tab:good_runs} lists 
the runs used for alignment studies in this period.  
Events from run 2102355, a 100~GeV pion beam run without a B-field, were used as input to all 
algorithms for the production of alignment corrections. For the \Local approach, two other 
pion runs were used in addition.  
Further event selection details are given in Section~\ref{sec:approaches}. 

\begin{table}[hbpt]
\begin{center}
\caption{List of selected runs used to assess the alignment results.} 
\vspace{10pt}
\begin{tabular}{|c|c|c|c|}
\hline
 Run Number & Particle Type & Energy (GeV) & B field \\ \hline
 2102355 & $\pi$ & 100 & Off \\ \hline
 2102439 & $e$ &  20 & On\\
  2102400 &  $e$ &  50 & On\\
  2102452 &  $e$ &  80 & On\\
  2102399 &  $e$ & 100 & On\\
  2102463 &  $e$ & 180 & On\\ \hline
  2102442 &  $\pi$ & 20 & On\\
   2102365 &  $\pi$ & 100 & On \\ \hline
\end{tabular}
\label{tab:good_runs}
\end{center}
\end{table}
%
  
\subsection{Simulation \label{sec:simulation}}

The CTB setup was simulated with Geant 4 using 
the same geometry description as the event reconstruction.
Detector positions and initial numbers were provided through 
an Oracle-based conditions database (look-up information) 
which allowed the five different periods to be distinguished
from one another.

CTB specific modifications were applied to the simulation  
for studying the Pixel and SCT alignment, i.e. the propagation through material 
upstream of the ID and the inclusion of measured beam profiles. 
The upstream material (mainly air and triggering/monitoring
scintillators) corresponded to 13.2\% radiation lengths and was taken into 
account to mimic the momentum distribution in the data properly. 
Profiles, consisting of beam incidence positions and angles, were taken 
from the data and were applied during the upstream simulation to bring the 
simulated hit maps and residual distributions of the silicon modules into agreement with the data.

The magnetic field map was calculated taking into
account the magnet geometry, in one quadrant of the transverse plane
with respect to the beam axis. 
The remaining field map was modeled assuming a symmetric
the field map around the main axis
of the magnet. The field map calculated along these lines compares
well with the actual measurement of the dipole field which were
performed before and after the CTB runs.

\subsection{Tracking and Reconstruction\label{sec:tracking}}

The default tracking algorithm in the CTB was the `CTBTracking' algorithm \cite{Thijs_phd}. 
CTBTracking consists of a pattern recognition part, developed specially 
for the CTB, and a track fitting
algorithm that is in use in full ATLAS as well as in the CTB. The pattern 
recognition finds the tracks by looping through combinations of 
space points. The track fitting algorithm is
based on a global $\chi^2$ minimization technique, often called the 
`breakpoint' method in the literature\cite{breakpoint_ref}. Multiple scattering and energy loss 
enter into the algorithm as additional fit
parameters at a given number of scattering planes. The track fit has a 
custom description of the detector material in the test beam setup, with 
one scattering plane for each layer of
silicon modules. This material description was precisely tuned to give the 
best possible track resolutions, down to very low energies (1 GeV or 
  less). A number of options and
features exist in the track fit that are particularly useful for the 
alignment algorithms, such as the possibility of setting the momentum to a 
fixed value in the fit, and the ability to
retrieve the fitted scattering angles and their covariances.

\begin{figure}[tbp]
\begin{center}
\includegraphics[width=0.85\textwidth]{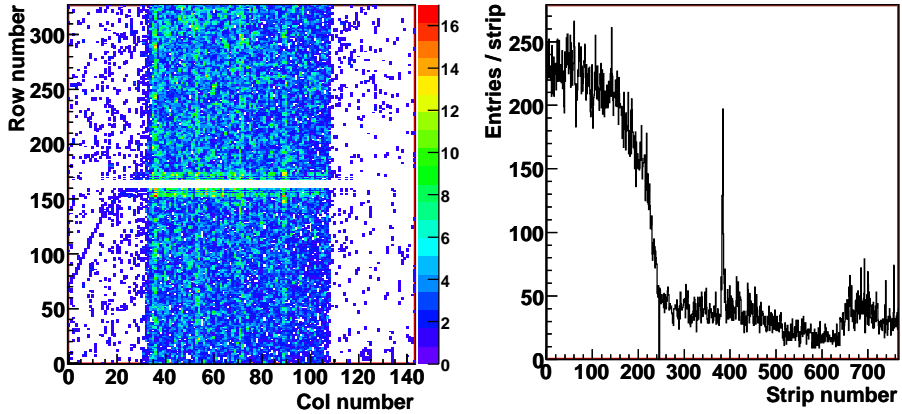}
\caption{Left: Pixel hit map for 100 GeV pion runs. 
The $x$-axis corresponds to Pixel $\eta$-coordinate and the $y$-axis to
  Pixel $\phi$-coordinate.  Empty horizontal bands correspond to the ganged
  pixels.  Lighter vertical bands are due to the 600-$\mu$m-wide
  pixels. Right: SCT hit map for the same run.}
\label{fig:HitPix}  
\end{center}
\end{figure} 

Fig.~\ref{fig:HitPix} shows typical hit maps for a Pixel and SCT
module.  The illumination was rather uniform for the channels that
lay within the scintillator trigger acceptance window in the central region ($\sim 3\times
3~\mbox{cm}^2$ wide). More details on the tracking performance of the
pixel detectors can be found elsewhere~\cite{CTB_Pixel}. 
Unmasked noisy channels can be distinguished in the SCT hitmap. 
Those that were masked during data-acquisition
appear as zero-entry channels.  The illumination was not uniform and limited 
along the strip length but only in the central region, 
where the sensor planes overlapped completely with the trigger scintillator.  

The limited illumination of the sensors had direct consequences on some of
the alignment degrees of freedom (DoF) due to insufficient constraints and
reduced sensitivity.  The problem was more severe for SCT modules, 
because the SCT modules were not tilted with respect 
to the beamline.  As the beam incidence was almost
perpendicular to the module planes, the alignment procedures were not very sensitive 
towards misalignments along the beam axis. 

The pixel sensors require free space in order to bond the readout chips
on the surface of the sensor.
In the precise $\phi$ coordinate,
unbonded pixels are physically connected to
nearby pixels (ganged pixels) and share a readout logic channel.
Due to this connection, whenever a hit was registered by a logic
channel, there was an ambiguity as to which pixel fired.  
In the long coordinate wider pixels (600-$\mu$m
instead of 400-$\mu$m) are used. The wider pixels
collect more hits. The impact of both effects is clearly seen
in the pixel module hit map (Fig.~\ref{fig:HitPix}). 
The ambiguity in the ganged pixels was also found to effect the alignment.
In a highly misaligned environment, tracking may make too many wrong decisions between 
ganged pixels. It was found that, in the presence of a high track quality cut, the ganged pixel hits
were favoured, degrading the quality of the alignment~\cite{robustalignalg}.

\begin{figure}[tbp]
\begin{center}
 \includegraphics[width=0.95\textwidth,angle=0]{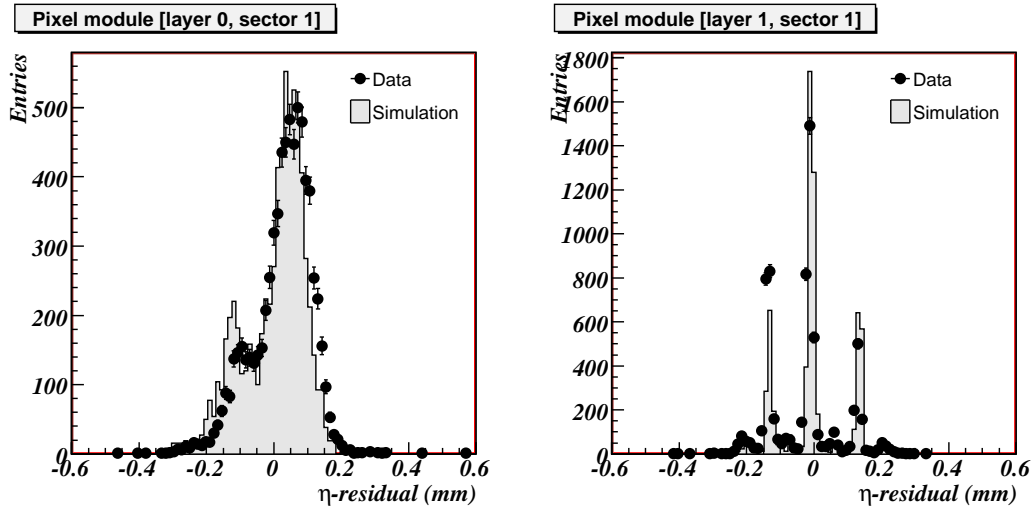}
 \caption{Pixel $\eta$-residuals for  tracks reconstructed with Pixels
 and  SCT.   Left:  first  Pixel  layer.  Right:  middle  Pixel  layer.
 \label{fig:PixEtaRes}}
\end{center}
\end{figure}
      
The fact that the modules were exposed to almost perpendicular beams
resulted in discrete
Pixel $\eta$-residual distributions.
Due to the large dimension in this direction (400 $\mu m$ compared to 300
$\mu m$ of the thickness of the silicon bulk) the drift of the charge
carriers along that direction is negligible.  Therefore, almost
all of the clusters consist of a single pixel in the
$\eta$-coordinate.  As the cluster position is located in its
geometrical center, the outcome is a discrete positioning of clusters
(Fig.~\ref{fig:PixEtaRes}).  With only three pixel layers providing
three precision points, a discrete residual distribution was obtained.
The use of SCT clusters in the tracking partially removed this
undesired effect~\cite{Sergio_phd}.  Effectively the pixel
$\eta$-residuals of the first and last pixel layers were somewhat broadened 
by overlaps of Gaussian distributions,
while the middle layer $\eta$-residuals remained discrete.
This peculiarity of the CTB setup made the alignment along the pixel
$\eta$-coordinate difficult. 
ATLAS collision data will not present such difficulties.

%
%

\section{Alignment of the CTB Data \label{sec:approaches}}
The goal of alignment is to determine the corrections
to the parameters that describe the position and orientation of the module in space. 
Each module is treated as a flat rigid body with 6 DoFs,
{\em i.e.}, three translations along the local
coordinate axes ($x$, $y$, $z$) and three rotations ($\alpha$, $\beta$, $\gamma$)
around the local coordinate axes, in a
right-handed orthogonal frame where the 
origin is at the center-of-gravity of each module and the
local $x$-coordinate is along the most precise coordinate. 
The translations correspond to the shift of the module with respect to its nominal
position.  For the axes orientation, the Cardano
representation of angular rotation with respect to the cartesian axes
was used. The alignment corrections were stored in the conditions database.

The alignment corrections are given in terms of CLHEP \cite{CLHEP} transform 
objects $H$, made of a rotation matrix $R$ and a translation vector $\vec{T}$. 
The rotation matrix is defined as:
\begin{equation} 
R=R_x(\alpha)\cdot R_y(\beta)\cdot R_z(\gamma) 
\end{equation} 
with $\alpha$, $\beta$ and $\gamma$ being the rotation
angles around the $x$, $y$ and $z$-axes. $R_z(\gamma)$ is the first
rotation applied and $R_x(\alpha)$ the last.  
The representation of a point in the local reference frame ($\vec{P}$) of a
module is $ H\cdot\vec{P}=\vec{T}+R\cdot\vec{P} $ in
the global frame.  Lets consider $H_0$ as the transformation
specifying the nominal position of a given module.  If $\delta H$ is a
shift of the module, the new transformation of the points measured by
it becomes $H = H_0\cdot\delta H$. Therefore, the task of the alignment is
to determine the 6 DoFs that define $\delta H$ for each module.
In the case
of poorly constrained movements, some DoFs may not be
considered. 

The technique to align each silicon module consists of  minimizing its two
residuals (pixel modules measure two coordinates and each SCT module has two
sensor  planes).  The  $i$-$th$  residual (defined  by  coordinate, plane  or
module)  is thus  $r_i  \equiv(\vec{m_i} -  \vec{e_i}(\mathbf a,  \pib))\cdot
\hat{k_i}$, where $\vec{m_i}$ represents the  position of the hit recorded in
the sensor plane, $\vec{e_i}$ is the intersection point of the extrapolated
track to the detector that depends on the alignment parameters $(\mathbf a)$
and the vector of track parameters $\pib = (d_0, \phi_0, z_0,
\theta_0,q/p)$.  $\hat{k_i}$ denotes  the unit  vector of  the measurement
direction
\cite{globalX2_note}. 

All alignment algorithms were run iteratively.
Initially, nominal detector and hit positions were used for
track reconstruction. After the track fit, residuals and
their derivatives with respect to alignment and/or track parameters 
were calculated to determine the alignment corrections. 
For each module, the best fit estimates for alignment parameters were derived and
its position was updated. A new reconstruction with updated module
positions was performed and the alignment was reiterated. This procedure is expected to
converge to final alignment corrections for each module and the residual resolution is expected to improve.

The alignment was performed using two different classes of approach. 
The \Robust approach is based on iterative minimization of the residual means 
of overlapping and non-overlapping modules. 
The approach is ``robust" because the output is stable against 
changes in the input tracking information. 
  
The \Valencia, \Local and \Global approaches are 
based on the linear least squares minimization defined for a set of 
reconstructed tracks as:
\begin{equation} 
\label{eqn:BigChi2Definition}
  \chi^{2}(\mathbf a, \pib_{1}, \dots, \pib_{t}) = \sum_{i\ \in\ tracks}
{\mathbf r_{i} \, ^{T} V_{i}^{-1}\mathbf r_{i}}
\end{equation} 
where $\mathbf r_{i} = \mathbf r_{i}(\mathbf a,
\pib_{i})$ is the vector of residuals measured for the fitted
track $i$. 
$V_{i}$ is the covariance matrix of the
residual measurements of \mbox{track $i$}.
The generic solution for alignment corrections ($\delta \mathbf a$)
is:
\begin{equation}
  \delta \mathbf a 
  = - \left(\sum_{i\ \in\ tracks} \left( \frac{d \mathbf{r}_i}{d \mathbf a}\right)^T V^{-1}
   \left(\frac{\partial \mathbf r_i}{\partial \mathbf a}\right)\right)^{-1}
  \!\!\!\!\!\!\!\! \sum_{i\ \in\ tracks} \left(\frac{d \mathbf r_i}{d \mathbf a}\right)^T V^{-1} {\mathbf r}_i
\quad \!\!\!\! = - A^{-1}  \quad \!\!\!\!\!\!\! \sum_{i\ \in\ tracks} \left(\frac{d \mathbf r_i}{d \mathbf a}\right)^T V^{-1} {\mathbf r}_i ~~,
\label{eq:global_approach_2}
\end{equation}
where $A^{-1}$ is the covariance matrix for $\delta \mathbf a$.
The size and contents of the matrix $A$ 
depend on the details of the alignment method which  are explained in
the following sections.

\subsection{The \Robust approach \label{sec:robust}}
The Robust alignment approach \cite{robustalignalg} is an iterative method to align 
a silicon detector with overlapping modules. In each
iteration alignment corrections are calculated from  
measurements of mean residuals, $\overline{res}$, 
and mean overlap residuals, $\overline{ovres}$, in the $x$ and $y$ 
coordinates. Overlap residuals are defined as the difference between two residuals from
two overlapping modules.
$y$ SCT residuals are
constructed using both hits from each side in a module. The
algorithm only corrects for shifts in 
the plane of the module. The alignment corrections
are given by:
\begin{equation}
  \label{eq:robust1}
a_{x/y} = - \sum_{j=1}^{3} \frac{s_j}{(\delta
s_j)^2} / \sum_{j=1}^{3} \frac{1}{(\delta s_j)^2} ~.
\end{equation}
$s_1$ to $s_3$ are defined as:
$s_1 = \overline{res}; s_2 = \sum \overline{ovres}_{x}; s_3 = \sum \overline{ovres}_{y}$, 
 where $\delta s_j$ are the measurement uncertainties. 
The range of the sum depends on the geometry of the detector. Given the simple CTB
geometry, a straightforward 
implementation of Eqn.~\ref{eq:robust1} was used. 

The alignment corrections were obtained as follows:  
The fractions of non-overlap and overlap hits in the sample
were controlled by coefficients $A$ for 
overlap hits  and $B$ for non-overlap hits, 
to adjust the influence of each information on the $x$ and $y$ correction. 
The corrections were weighted with the ratio of the 
total number of overlap hits $A \cdot noh_{x/y}$ and the number of hits $B \cdot nh_{x/y}$
to the total number of hits,  $N_{x/y}$. 
The total residual weight $rw_{x/y}$ and the total overlap residual weight
$orw_{x/y}$ obtained this way corresponded to $1 / \delta {s_j}^2$: 
\begin{equation}
  rw_{x/y}  =  \frac{B \cdot nh_{x/y}}{N_{x/y}} ~, \quad 
  orw_{x/y}  =  \frac{A \cdot noh_{X/y}}{N_{x/y}}
 \label{eq:weights}
\end{equation}
There was one overlap for
each two modules in a layer. Thus, this information could be used for only one
sector which was arbitrarily chosen to be sector 1. The alignment
corrections for modules in Sector 1 is given by Eqn.~\ref{eq:alignment1} 
and in Sector 0 is given by Eqn.~\ref{eq:alignment0}, as: 
\begin{equation}
  a_{x/y}   =  - orw_{x/Y} \cdot \overline{ovres}_{x/y} - rw_{x/y} \cdot \overline{res}_{x/y}  
\label{eq:alignment1}
\end{equation}
\begin{equation}
  a_{x/y}  =  - \overline{r}_{x/y} 
 \label{eq:alignment0}
\end{equation}


The CTB alignment was carried out using ``unbiased" residuals,
{\em i.e.}, the hit of the aligned wafer on the side of the module
was removed from the track fit.  
About 72,000 events from run 2102355 were used for the alignment. 
This run contained about 10 to 50 times more hits than overlap hits. 
Information from residual distributions and overlap
residual distributions were weighted so that overlap residuals had
almost similar influence: setting A to 10 and B to 1 was found optimal.
Further tests showed that other values affected
the speed of convergence rather than the final
result.

There were two major limitations in the application of the \Robust algorithm to the CTB data.
First, 
significant tilts arose from the hand-mounted modules in the setup.
In contrast with the other algorithms, the
\Robust algorithm does not correct for rotations. Therefore, after
alignment, the residuals still had a global $Y$ dependence, 
in agreement with the tilts observed 
around the Pixel $y$-axis (see Sec.~\ref{sec:globalx2}).
The dependence vanished when the modules were rotated
accordingly. This is the main reason why the residual resolution
after the \Robust alignment were not as 
good as the ones achieved by other algorithms. The modules with the
largest residuals after the \Robust Alignment correspond to the modules
with the largest rotations. 
Second, discrete Pixel $y$ ($\eta$) residuals 
resulted in less stable mean of 
the residuals with respect to any small shifts. 

\begin{figure}[tbp]
\begin{center}
\hspace{-2cm}
\includegraphics[width=0.5\textwidth]{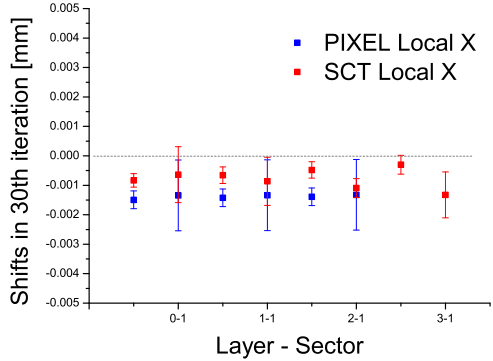}
\vspace{-0.5cm}
\caption{\it Shifts in the $x$ direction for the 30th iteration of the \Robust 
alignment. Statistical uncertainties
on the measurements are represented by the error bars. 
The quality of the reconstructed track parameters is insensitive to this negligible systematic effect. 
} 
\label{fig:CTB30thIter}
\end{center}
\end{figure}

The \Robust algorithm converged on a solution without a tight 
track selection. 
Although 30 iterations were performed to align the detector,
stable results were achieved after 15 iterations. 
The residuals improved significantly and the 
track quality stabilized after a small 
number of iterations.
After 30 iterations, 
about 1~$\mu$m global shifts of module positions in the negative $x$ direction were observed
(Fig.~\ref{fig:CTB30thIter}).
The \Robust algorithm had the advantage of requiring minimal computing resources.
The CPU time used by the algorithm were shown to be negligible compared to that of the preceding 
track reconstruction.

\subsection{The \Valencia approach \label{sec:valencia}}
The \Valencia alignment algorithm \cite{Sergio_phd} is based on the
numerical minimization of the $\chi^2$ function defined in
Eqn.~\ref{eqn:BigChi2Definition} using ``biased" residuals ({\em i.e.}, the
hit of the module being aligned is included in the track fit).
The covariance matrix is assumed to be diagonal
and the diagonal elements are filled with the measurement uncertainties, 
$\sigma_{r_i}$, for residuals, $r_i$,
both of which are calculated numerically.  The
only fit parameters are the alignment corrections, ignoring the
correlation between track and alignment parameters.  The algorithm is
therefore executed iteratively, alternating between track and alignment fits.

\begin{figure}[hbt]
\begin{center}
\includegraphics[width=0.55\textwidth]{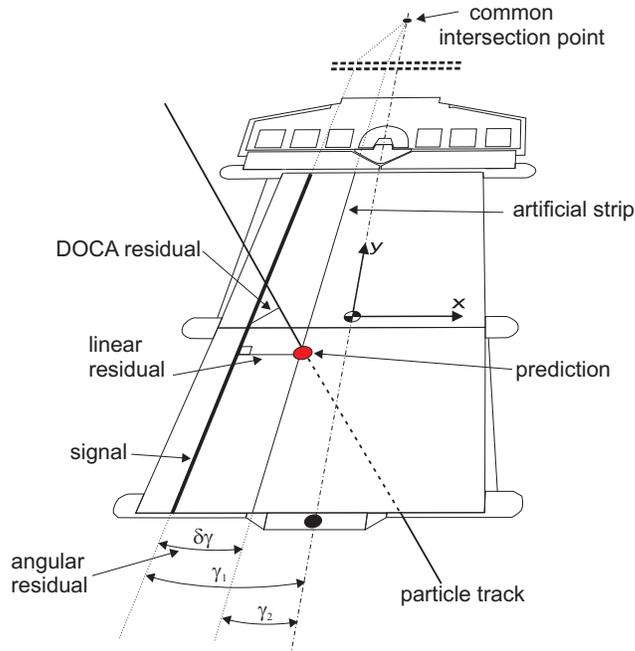}
\caption{Residual definitions in a SCT end-cap module used by the different 
alignment algorithms: the angular and linear residuals and distance of closest approach (DOCA).}
\label{fig:angular_res}
\end{center}
\end{figure}

The SCT endcap outer module strips follow a fan-out geometry and thus have
a variable pitch along the vertical direction (Sec.~\ref{sec:setup_data}). 
Therefore, instead of using the standard
{\em ``linear''} residual (perpendicular distance from the track prediction to the
strip), {\em ``angular''} residuals ($\delta\gamma$) were used (Fig.~\ref{fig:angular_res}). 
These represent the difference 
between the angular separation of the signal channel and a ``fictitious'' 
strip passing through the extrapolated point. The strip-pitch dependence 
was thus avoided, and uniform angular residuals were obtained. 

The outlier hit rejection was applied by defining an acceptance region
determined by a critical value of the $\chi^2$ (outlier rejection).
This value was taken as three standard deviations with respect to the
mean value of the reduced residual distribution $(r_i/\sigma_{r_i})$
calculated before the minimization.  The fraction of measurements
lying out of the acceptance region was $\sim $3\%, and reduced to
below 1\% if five standard deviations were used.

\begin{figure}[tbp]
\begin{center}
\includegraphics[width=0.50\textwidth]{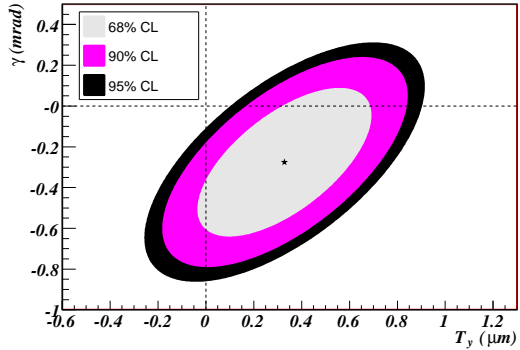}
\caption{Contour ellipses of constant probability content for three
different confidence levels and two alignment parameters of a Pixel module; translation along
the most sensitive coordinate (horizontal axis) and tilt angle (vertical axis), as obtained 
by the \Valencia alignment algorithm.}
\label{fig:ellipses}
\end{center}
\end{figure}

The \Valencia algorithm was intended for runs without magnetic field
yielding straight tracks.  After reconstruction, each track was
extrapolated to the silicon modules. If the extrapolation lay outside
the module geometrical acceptance, the track prediction was
discarded. The module intersection point of the accepted tracks was
transformed into the local frame and residuals were calculated. For
Pixels, only measurements in the $\phi$ ($x$) direction were
considered. The $\eta$-coordinate was ignored due to the non-Gaussian
residual distributions (Fig.~\ref{fig:PixEtaRes}).  For the SCT modules, angular residuals
and measurements from both SCT sides were
used\footnote{Except for module [layer 2, phi 1] with a single working
plane.}.
Although an analytical residual linearisation as a function of the alignment parameters was
not computed, the dependence of the $\chi^2$ on the alignment parameters
remained linear. Fig.~\ref{fig:ellipses} shows the contour regions for two fitted
variables and three different confidence level intervals (68\%, 90\% and
95\%) for one Pixel module.

The alignment was performed in three consecutive steps, with variable
number of iterations in each step: ({\em st-1}) internal alignment of
the Pixel modules, ({\em st-2}) broad alignment of the SCT modules
with respect to the Pixel system, and ({\em st-3}) fine alignment of
all silicon modules.  In {\em st-1} ($\sim$ 6 iterations), tracks
lying in the overlap region between Pixel modules in the same layer
were selected to enhance the number of overlap hits and to produce 
a pixel alignment.  In {\em st-2} ($\sim$ 2 iterations), tracks
reconstructed only with the pixel hits were extrapolated to the SCT
planes.  In this manner, it was possible to compute SCT residuals
(unbiased only in this case) which served as input for 
an initial alignment of the SCT modules with respect to the Pixel modules.
The required correction
of the SCT modules was several hundreds of microns. In {\em
st-3} ($\sim$ 8 iterations), all silicon modules were included in the
track fit and all were aligned simultaneously. In
this last stage the alignment corrections per module were of few
micrometers.

During alignment, the first Pixel module [layer 0, phi 0] was kept as
an anchor; fixed to its nominal position to fix the global
degrees of freedom.  DoFs to which the sensitivity was
very small were excluded from the set of fitted alignment parameters.
Module positions along the beam axis were not considered. 
For Pixels, only the displacements along the sensitive
coordinate were fitted.  The tilt angle ($\gamma$) was excluded
in ({\em st-1} and {\em st-2}), but fitted in step ({\em st-3}).  For
the SCT modules, the parameters for displacements along and across the
sensitive coordinate together with the in-plane rotation were fitted
in all steps. The inclusion of one
additional angle ($\beta$) during the last iterations was found to marginally
help to improve the results for both sub-detectors.

\subsection{The \Local approach \label{sec:localx2}}
The \Local approach \cite{MPP-2005-174, MPP-2006-118} derives from
Eqn.~\ref{eqn:BigChi2Definition}.
The $\chi^{2}$-function 
uses unbiased residuals, which are defined
as the 3D distance of closest approach (Fig.~\ref{fig:angular_res}).  
The algorithm uses a diagonal covariance matrix, $V$ 
that is simlar to that of the \Valencia approach.  
The residual errors are calculated using hit errors
and the extrapolated tracking errors. 

The \Local algorithm produces alignment constants for each module
separately, neglecting correlations between the modules during an iteration. 
Thus, the solution reduces to inverting as many $N\times N$ matrices as there are
modules, where $N$ corresponds to the DoFs of each module (up to 6).
Track parameters with a better fit quality gradually bring correlations into play
after every iteration.

The fact that CTB was found to be a degenerate setup for track-based alignment
required inclusion of external
constraints to resolve some of the degeneracies. 
These were a momentum constraint to the reconstructed tracks and an  
additional stabilization term to the diagonal elements of the matrix  
$A$ in Eqn.~\ref{eq:global_approach_2}.  The stabilization term acts like an additional  
measurement with a zero residual, full sensitivity in the  
corresponding degree of freedom (the derivative in Eqn.~\ref{eq:global_approach_2} is equal to  
one) and an uncertainty $\sigma_{\mathrm{stability}}$. The  
uncertainty  $\sigma_{\mathrm{stability}}$ corresponds to the inverse  
of the square root of the added term. These additional stability  
terms constrain the movement to be within $\sigma_{\mathrm{stability}} 
$. The values for $\sigma_{\mathrm{stability}}$ are 10, 10, 100 $\mu$m for  
the Pixel $x$,$y$,$z$ coordinates and 100 $\mu$m for the SCT $x$,$y$,$z$ coordinates. 
For the module  rotations the value for $\sigma_{\mathrm 
{stability}}$ is set to one $mrad$.

The momentum of the incident particles from SPS is known more
precisely than the intrinsic momentum resolution of the CTB ID
setup. Consequently, this information can
be used to constrain the track curvature. 
Tracks with different beam energies were used as input, using 10,000 events from each pion run 
listed in Table~\ref{tab:good_runs}. The alignement procedure 
was parallelized where multiple jobs with different momentum
constraint settings were executed simultaneously.
When jobs were finished the alignment information was
collected and merged.  Subsequently a new iteration with a new set of parallel jobs was started.

The usage of overlap hits, a hit
lying in the overlap region of two modules on the same layer, has
a profound impact on alignment by 
constraining relative positions
of both sectors, thus avoiding divergences due to lack of external constraints.
Residual calculation is also more precise for overlap hits than non-overlap
hits.
Non-overlap hits were rejected for
alignment once a defined limit was reached.  In this way the number of
overlap hits was enriched with respect to the number of non-overlap
hits. The maximum number of
non-overlap hits was set to 400.

\begin{figure}[tbp]
  \begin{center}
    \includegraphics[width=0.98\textwidth]{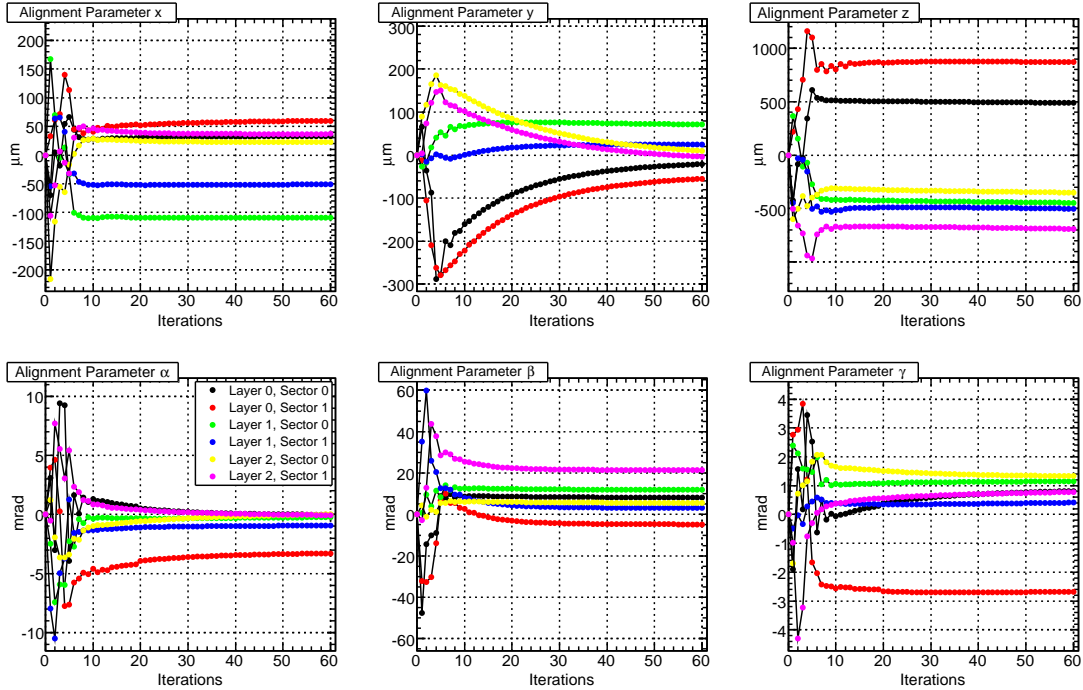}
    \caption{Flow of alignment parameters of the 6 Pixel modules through the iterations of the 
\Local alignment algorithm.
     \label{fig:localChi2PixelFlow}}
  \end{center}
\end{figure}

For the alignment the iteration chain was performed 60 times. The
flow of the 6 alignment parameters of each Pixel module through the
iterations is shown in Fig.~\ref{fig:localChi2PixelFlow}.  
After 10 iterations, nearly all degrees
of freedom of all modules converged on stable values. 
Slower convergence of some parameters was due to the imposed
stability term.  The procedure was stopped after 60 iterations,
when no significant improvement of track parameters was observed
and alignment corrections for the sensitive coordinates were at the
submicrometer level.

\subsection{The \Global approach \label{sec:globalx2}}
\label{sec:global_approach}
The \Global algorithm \cite{globalX2_note,globalX2_mumbai} is based on the minimization
of the $\chi^2$ defined as Eqn.~\ref{eqn:BigChi2Definition}
with respect to the alignment parameters. 
The residuals are defined within the module plane and are biased ({\em i.e.}, the
hit of the module being aligned is included in the track fit). They
depend on the track parameters (\pib) as well as on the subset of
alignment parameters related to the intersected module ($\mathbf a$):
\begin{equation}
\frac{d \mathbf r}{d \mathbf a} = \frac{\partial \mathbf r}{\partial \mathbf a} + \frac{\partial \mathbf %
r}{\partial \pib} \frac{d \pib}{d \mathbf a} ~.
\label{eq:global_approach_3}
\end{equation}

The method has the advantage of properly treating all correlations
between residuals arising from common track parameters and Multiple
Coulomb Scattering (MCS). Solution~\ref{eq:global_approach_2} requires
inverting a symmetric matrix of size $N\times N$, where $N$ is the
number of DoFs of the problem. For large systems
(for instance, the entire ATLAS ID), the solution with
accurate numerical precision and in a reasonable CPU time could be a
challenge \cite{scalapack_mumbai}. 
In the CTB case, however, the system consisted of just 14
silicon modules. Therefore it was free from such numerical limitations. 
Intrinsic alignment of an unconstrained system always leads to a singular matrix 
and consequently an ill-defined solution.  This is best solved by
diagonalization of the matrix.  The singular modes can subsequently
be ignored in the solution.  The procedure can be further extended to
remove all ``weak modes'' which either represent unphysical
deformations or have an associated error exceeding expected
misalignments.

In order to solve the CTB alignment, the following approach was
adopted: two anchor modules were chosen (the first Pixel and the last
SCT) which removed the exact singularities from the solution.
All considered tracks were nearly
parallel to one another and orthogonal to the SCT module planes. Also the $y$ tilt angles of  the 
Pixel modules were considered to be very accurately known from the survey.
Consequently the following DoFs were
removed from the fit: out of the plane translation and the rotations with
respect to $x$ and $y$-axes. This choice resulted in 3 DoFs per module
(36 in total).
However, results indicated a substantial residual misalignment related to the uncorrected  $y$ rotation 
of the Pixel modules. 
The largest misalignment was found for the upper module in layer 2 with a value of $25.2 \pm 0.5~mrad$. 
The $y$ rotations of the Pixel modules were eventually included in the alignment fit which 
efficiently eliminated the corresponding misalignments. 

\begin{figure}[tbp]
\begin{center}
\includegraphics[width=0.42\textwidth]{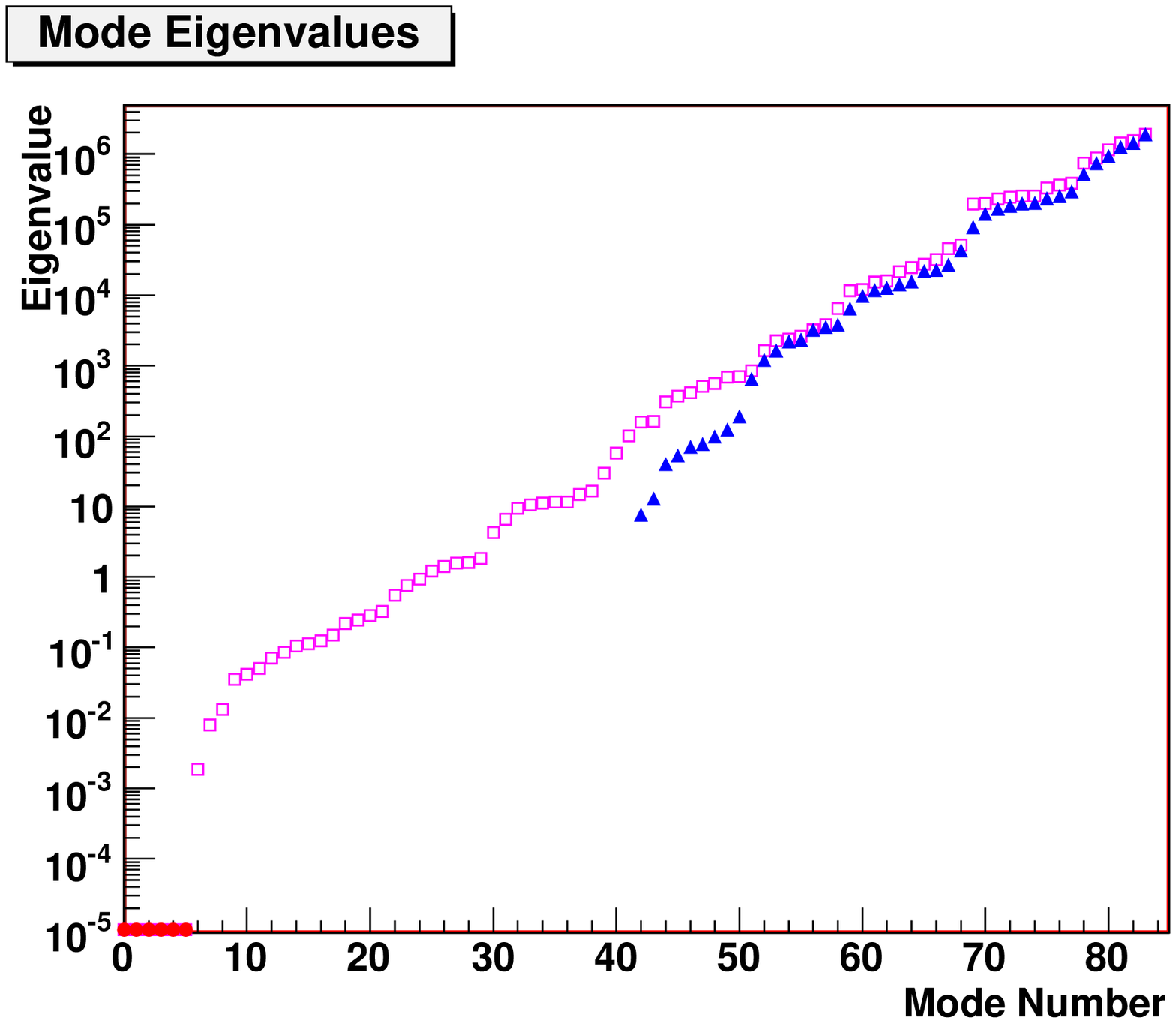}
\includegraphics[width=0.42\textwidth]{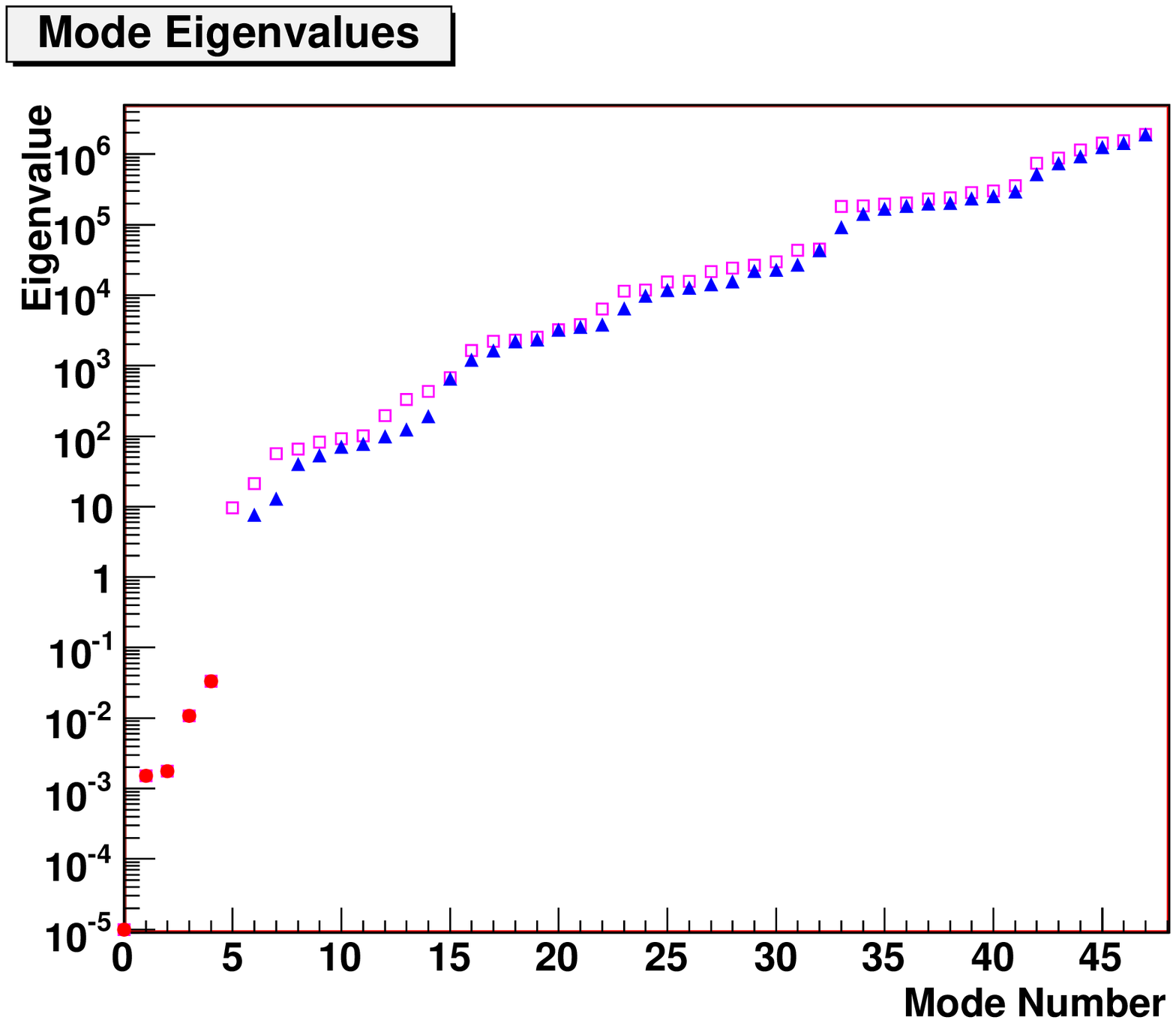}
\caption{Left: original eigenspectrum of the unconstrained solution (84 DoFs - open squares) 
overlaid with that of the actual solution (42 DoFs - solid triangles) of \Global algorithm.  Six
solid circles on the leftmost are the singular modes of the asymptotic freedom of the system. 
Their actual value is zero up to the numerical precision and were fixed to $10^{-5}$ for
clarity only. Right: actual solution eigenspectrum (42 DoFs - solid triangles) compared to the analogous
one without the anchor modules requirement (open squares).
The five solid circles are the new near-singularities.
\label{fig:global_eigenspectra}}
\end{center}
\end{figure}

The final alignment 
was concerned with 42 DoFs.  Fig.~\ref{fig:global_eigenspectra} shows the
comparison of eigenspectra (obtained by {\em DSPEV} routine from the
LAPACK library~\cite{LAPACK}) of the unconstrained CTB geometry and
the one used for the final alignment.  Elimination of unphysical
parameters efficiently removed the lowest part of the eigenspectrum.
Fig.~\ref{fig:global_eigenspectra} gives also the comparison of the final
alignment to the one without anchor modules. The five weak modes
correspond to the approximate\footnote{Axes of local reference systems
in different modules are not parallel which lifts the perfect
translational degeneracy. Similarly, rotations are free only
approximately.} freedom
of two global translations and three rotations of the entire setup.

The method required four iterations for convergence, however a total of seven iterations was used 
on about 50,000 events at each iteration. 
Translations of some SCT modules in $y$-direction were found to be as large as 1.5~mm;
$x$ translations never exceeded 0.4~mm.

%

%
%
\section{Results}

In order to assess the quality of the alignment, one must check the
track reconstruction quality and physics observables.
For this purpose, the alignment
corrections were applied to the data detailed in Table~\ref{tab:good_runs}.  
 


After aligning the modules, the track finding efficiency increased. For example,
for the \Robust alignment approach, the number of tracks per event was found to stabilize 
at around 0.95. As expected, an average of three hits in
the pixels and eight in the SCT (two per module) were found.
All four alignment approaches produced similar performances, consistent with the 
simulation.

\begin{figure}[tbp]
\begin{center}
\includegraphics[width=0.90\textwidth,angle=0]{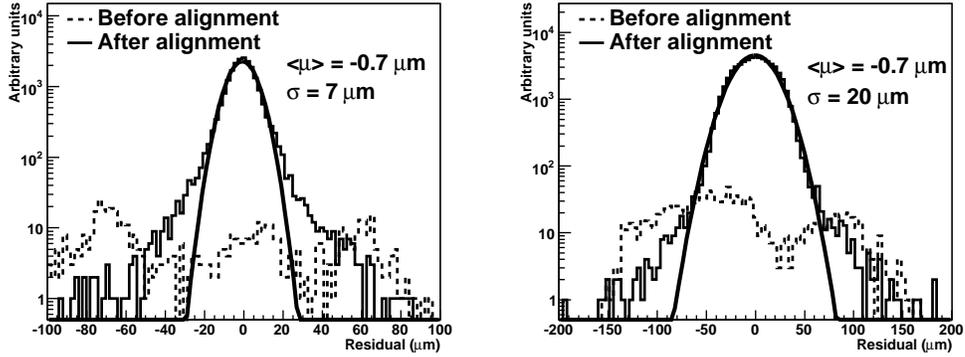}
\caption{Residuals of all Pixels (left) and all SCT (right) modules 
  before and after alignment. The width of the distribution is consistent with the intrinsic resolution of the modules.}
\label{fig:residuals}
\end{center}
\end{figure}

\begin{figure}[tbp]
\begin{center}
\includegraphics[width=0.8\textwidth,angle=0]{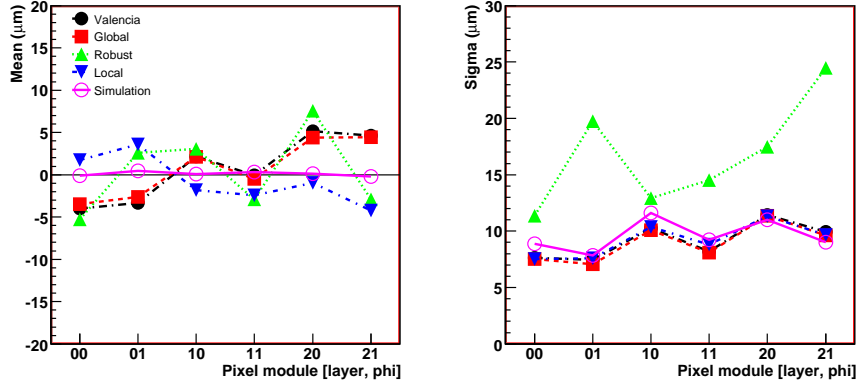}
\caption{Pixel biased $\phi$-residuals mean (left) and $\sigma$
(right) distributions for the aligned data and the 
simulation, for the 20 GeV pion run with B=1.4 T.\label{fig:pix_res_mean_comp}}
\end{center}
\end{figure}

A well-aligned setup returns residuals with a mean of zero and a width 
consistent with the intrinsic resolution of the detector and the track fit errors. 
Fig.~\ref{fig:residuals} shows the biased $x$-residuals of all the Pixel and SCT
modules for the 100~GeV pion run, for those tracks which had at least three pixel and six SCT hits. The  width of the
distribution after alignment is consistent with  the intrinsic  resolution of
Pixels and SCT  modules.  Fig.~\ref{fig:pix_res_mean_comp}  shows the
mean of Pixel module residuals for an example run (20 GeV/c pion
run). While simulation residual means are centered around zero for all modules, 
the aligned detector data show fluctuations.   
From the size of the fluctuations, we conclude that the Pixel residuals of all
alignment methods agree within $5~\mu m$ over the whole momentum range.
Fig.~\ref{fig:pix_res_mean_comp} also shows a good agreement between the $\chi^2$ 
minimization methods and the simulation on the residual resolutions.
The \Robust method results in a worse residual resolution since this method only corrects for
alignment  shifts in the module  plane.
Fig.~\ref{fig:pix_res_mean_comp}  also  reveals a
dependence  of the  $\sigma$  of  the pixel  residuals  on the  module
number.   This  indicates contributions  to  the  resolution from  the
geometry of the setup in addition to the intrinsic detector resolution.
The residuals also vary because the track error varies along the track due to MCS, for example.

\begin{figure}[tbp]
\begin{center}
\includegraphics[width=0.8\textwidth,angle=0]{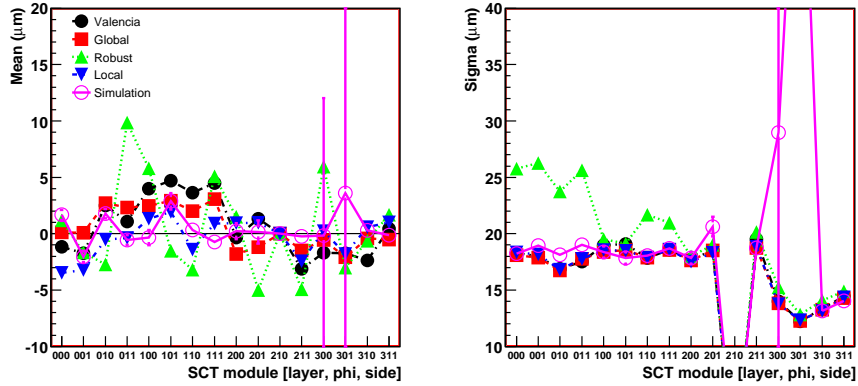}
\caption{SCT biased residuals mean (left) and $\sigma$ (right) distributions for the
aligned data and the simulation, for a 20 GeV pion run
with B=1.4 T. A single SCT module has two entries
in each plot (e.g. 000 and 001 correspond, respectively, to the module front and back plane
at layer 0, sector 0). The dip at index 210 in the right plot corresponds to the plane which was not working.
\label{fig:sct_res_mean_comp}}
\end{center}
\end{figure}

The residual mean distributions for each SCT plane are shown
in Fig.~\ref{fig:sct_res_mean_comp}. 
Systematic correlations in the signs of the means is observed among 
the alignment algorithm results. 
Fig.~\ref{fig:sct_res_mean_comp} also shows that the residual resolution of the 
SCT modules in aligned data (except those reconstructed using \Robust method alignment corrections) are 
around 20 $\mu m$, which is in good agreement with the simulation.

All track parameters at the perigee ($d_0$, $z_0$, $\phi_0$ and
$\theta_0$ and the momentum) were examined when tracks were
reconstructed with the alignment corrections from the four algorithms.  
The values of spatial track
parameters were not exactly similar for tracks reconstructed with different
constants, however, they followed consistent trends for the runs studied.
The difference can be attributed to the insufficiently constrained global degrees of
freedom. 
The residuals and curvature, hence the track fit $\chi^2$ and the  $p_T$, are 
invariant under rigid body translations and 
rotations of the whole system.

The reconstructed $\phi_0$ and $\theta_0$ values
depend on the beam properties as well as the  module locations provided by the
algorithms. Therefore,  
the measured $\phi_0$ and $\theta_0$ in data were used to tune
the beam spread in the simulation and to evaluate the interaction 
length, $X_0$, upstream of the CTB setup.  
The track $\phi_0$ and {\it $\theta_0$} resolutions 
improved with increasing momentum, as expected with a reduced MCS for more energetic particles. 

\begin{figure}[tbp]
  \begin{center}
    \includegraphics[width=0.70\textwidth]{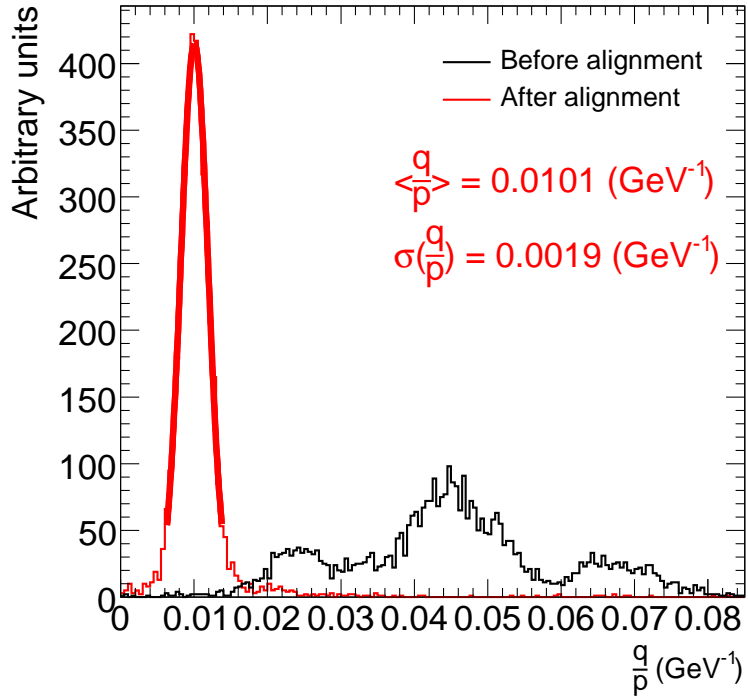}
    \caption{Momentum resolution of Pixel and SCT detectors for a 
                 100 GeV pion run with non-zero B-field with and without alignment corrections. 
             \label{fig:localChi2_MomentumCompare}}
  \end{center}
\end{figure}

\begin{figure}[tbp]
\begin{center}
\includegraphics[width=0.85\textwidth]{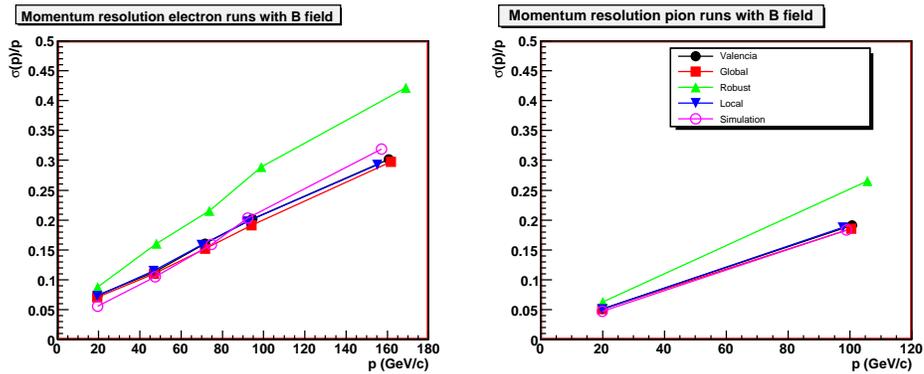}
\caption{Momentum resolution as a function of the reconstructed momentum.}
\label{fig:mom_res}
\end{center}
\end{figure}

The momentum reconstruction provides a very powerful test of the alignment performance.
Fig.~\ref{fig:localChi2_MomentumCompare} shows the recovery of the momentum resolution of the 
100~GeV pion run after alignment, from a highly degraded initial measurement. 
The momentum measurement does not depend on global transformations.  Therefore the
momenta of the tracks reconstructed with different alignment constants ought to 
agree. Fig. \ref{fig:mom_res} is used to compare 
the electron and pion momenta resolution 
as a function of the reconstructed momentum 
obtained from the four alignment methods to the simulation. 
The momenta reconstructed using all algorithms, in particular 
$\chi^2$ minimization ones, are consistent with the
simulation.
The \Robust method returns slightly worse results
since the alignment does not take rotations of the modules into account.

The reconstructed electron momentum is
significantly less than the nominal (set by the beamline), for both data and simulation.
The presence of several layers of upstream material 
can account for this effect, because  
the radiated energy of electrons before they enter the tracking volume was not recovered. 
As pions do not suffer as much from energy loss,
their reconstructed momenta agree much better with the nominal set by the SPS.



The convergence of alignment corrections per iteration and the improved
residual distributions presented are mandatory
but not sufficient to ensure the success of the alignment. Unfortunately, 
survey data of the CTB detector setup does not exist, therefore, a comparison with the
derived alignment sets was not possible.  However, a comparison of the position
and orientation of each detector element derived with the four algorithms
served as a means of validation.

When the alignment constants for the four algorithms were compared, 
the algorithms were observed to provided large corrections
(several hundred microns). 
The chosen alignment strategy
(fixing one or several modules as opposed to leaving the whole system free floating
 or constraining or removing some DoFs from the alignment)
has an impact on the solution of the global DoF of the system.
In order to compare the results of the different
algorithms, they need to be globally matched.  
Allowing a global offset for each 
alignment set was chosen to be the method of finding a best match of the alignment results. 
After having subtracted the global offsets between the geometries,  it was observed that 
a good agreement between the algorithms for the
most sensitive coordinates $x$, $y$ and 
$\gamma$ was reached.
Given the low sensitivity of the
alignment procedures to the alpha and beta rotations, the agreement
between the algorithms for these coordinates was only marginally improved.

\section{Summary and Conclusion}

Four independent algorithms were used to successfully align the 
setup formed by the silicon modules of the ATLAS Inner Detector tracker,
using data collected during the 2004 Combined Test Beam. 
The reconstructed track parameters and hit residual distributions 
were studied.  The performance
of the alignment algorithms was assessed by comparing with a 
simulation, in which all modules were at their nominal positions.
The simulation can be taken as a benchmark where all errors were regarded 
as only being due to the intrinsic resolution of the modules.

All alignment approaches yielded results for the reconstructed momentum
of electrons and pions that agreed with the simulation.  
Slightly worse momentum resolution was observed using the
\Robust algorithm.  This was understood and explained by the fact
that the algorithm was limited to re-alignment of the two in-plane translations only. 
The unresolved residual misalignments (e.g. in-plane rotations) unavoidably led 
to reduced track fit quality and consequently increased uncertainties on the 
reconstructed curvature.
For the remaining track perigee parameters, consistent
results were obtained with each method.

All four methods agree well on the residuals for all modules and planes, and
with the simulation.  The resolution of individual pixel modules is around
10 $\mu$m and the SCT around 20 $\mu$m.  Observed differences for the residual 
mean values remain below 5 $\mu$m.  We conclude that
the silicon modules of the ATLAS ID were aligned at the CTB with a
precision of 5~$\mu$m in their most precise coordinate.

The data collected at the ATLAS Combined Test Beam in 2004 served as an 
invaluable test bed for the Inner Detector alignment algorithms. 
For the first time ever, the readiness of the alignment algorithms was assessed with experimental data.
All algorithms
performed satisfactorily given the limitations inherent to the CTB geometry
and the beamline arrangement. The narrow tower of modules and almost parallel particle beams 
gave rise to undetermined degrees of freedom. These were successfully dealt with 
by the four algorithms, each in its own way, providing consistent and high quality
measurements of the test beam track parameters.

%
%
\acknowledgments
 
We are greatly indebted to the CERN accelerator departments and to the H8 beam-test facility for their efforts in building and operating the 
beamline, to our technical collaborators for their support in constructing and operating the Combined Test Beam components and to the IT 
department for their support for the computing infrastructure. We are grateful to all the funding agencies which supported generously 
the construction and the commissioning of the ATLAS detector components and the computing infrastructure.

We acknowledge the support of ARC and DEST, Australia; Bundesministerium f\"ur Wissenschaft und Forschung, Austria;  
CERN; Ministry of Education, Youth and Sports of the Czech Republic, Ministry of Industry and 
Trade of the Czech Republic, and Committee for Collaboration of the Czech Republic with CERN; Danish Natural Science Research Council; 
IN2P3-CNRS and Dapnia-CEA, France; BMBF, DESY, and MPG, Germany; 
INFN, Italy; MEXT, Japan; FOM and NWO, Netherlands; The Research Council of Norway; Ministry of Science and Higher Education, Poland; 
Ministry of Education and Science of the Russian Federation, Russian Federal 
Agency of Science and Innovations, and Russian Federal Agency of Atomic Energy; JINR;  
Ministerio de Educaci\'{o}n y Ciencia, Spain; State Secretariat 
for Education and Science, Swiss National Science Foundation, and Cantons of Bern and Geneva, Switzerland; National Science Council, Taiwan; 
The Science and Technology Facilities Council, United Kingdom; DOE and NSF, United States of America.

%
%
\bibliographystyle{atlasstylem}
\bibliography{Main_JINST_ATLAS_CTB_IDalign}

\end{document}